\documentclass[12pt]{article}
\usepackage{amssymb,amsmath,epsfig}
\allowdisplaybreaks

\begin{document}
\title{\bf The Influence of Modification of Gravity on the Dynamics of Radiating Spherical Fluids}

\author{Z. Yousaf$^1$ \thanks{zeeshan.math@pu.edu.pk}, Kazuharu Bamba$^2$
\thanks{bamba@sss.fukushima-u.ac.jp} and M. Zaeem ul Haq
Bhatti$^3$ \thanks{mzaeem.math@pu.edu.pk}\\
$^{1,3}$ Department of Mathematics, University of the Punjab,\\
Quaid-i-Azam Campus, Lahore-54590, Pakistan\\
$^2$ Division of Human Support System,\\ Faculty of Symbiotic Systems Science,\\ Fukushima University, Fukushima 960-1296, Japan}

\date{}

\maketitle
\begin{abstract}
We explore the evolutionary behaviors of compact objects in a
modified gravitational theory with the help of structure scalars.
Particularly, we consider the spherical geometry coupled with heat and
radiation emitting shearing viscous matter configurations. We construct structure scalars by splitting the Riemann tensor orthogonally in $f(R,T)$ gravity with and without constant $R$ and $T$ constraints,
where $R$ is the Ricci scalar and $T$ is the trace of the energy-momentum tensor. We investigate the influence of modification of gravity on the physical meaning of scalar functions for radiating spherical matter configurations. It is explicitly demonstrated that even in modified gravity, the evolutionary phases of relativistic stellar systems can be analyzed through the set of modified scalar functions.
\end{abstract}
{\bf Report number:} FU-PCG-08

\section{Introduction}

Gravitation is probably easily conceived elementary interaction that
one can experience in everyday life. Relativistic study is seen to
be the foundation of modern physics along with the quantum theory.
In the exploration of celestial weak gravitational interaction,
relativistic effects must be taken into account. There are lucid
examples of relativistic stellar systems, such as white dwarfs,
neutron stars, black holes in which these effects may have major
outcomes. As a matter of fact, in order to study these systems, it
becomes necessary to take observationally viable gravity theories.
Further, many interesting results coming from observational
ingredients of Supernovae Ia, cosmic microwave background (CMB)
radiation, etc. \cite{ya1} have made a great revolution in the field
of cosmology and gravitational physics thus opening a new research
platform. These observations and experiments reveal that currently,
there is an accelerating expansion in our cosmos. According to the
recent observational results obtained from, e.g., the Planck
satellite~\cite{ya2, Planck:2015xua, Ade:2015lrj}, the BICEP2
experiment~\cite{Ade:2014xna, Ade:2015tva, Array:2015xqh}, and the
Wilkinson Microwave anisotropy probe (WMAP)~\cite{Komatsu:2010fb,
Hinshaw:2012aka}, the energy fraction of the baryonic matter is only
5\%, while that of dark matter and dark energy are 27\% and 68\%,
respectively.

Introducing modified gravitational theories after generalizing the
Einstein-Hilbert (EH) action to explore the mystery of cosmic
accelerating expansion is a very popular approach among relativistic
astrophysicists. Nojiri and Odintsov \cite{ya3} explained that why
extension to Einstein gravity theories are attractive in exploring
the evolutionary mechanism of cosmic late acceleration. Extended
gravity theories involve $f(R)$, $f(T)$ etc., where $T$ is the torsion scalar in teleparallel gravity (for further reviews on
dark energy and modified gravity, see, for
instance,~\cite{R-NO-CF-CD}). The simplest extension of the
Einstein's theory is $f(R)$ theory obtained by replacing the Ricci
scalar with its arbitrary function in the EH action. This theory was
brought in after few years from the advent of the Einstein's relativity
to analyze possible alternatives \cite{ya4} and was then studied
occasionally by several researchers \cite{ya5} to renormalize
general relativity \cite{ya6} which requires higher curvature
dark source terms in the EH action. This theory attracted many relativistic
astrophysicists in the possible explanation of cosmic inflation due to the
quadratic Ricci scalar corrections \cite{ya7} in the EH action.

In the similar fashion, many other extended gravity theories has
been discussed, like, $f(G)$ in which $G$ is the Gauss-Bonnet
invariant. Other curvature amalgams have also been employed such as
$f(R,G)$ and
$f(R,R_{\alpha\beta}^{~~\alpha\beta},R_{\alpha\beta\gamma\delta}^{~~~~\alpha\beta\gamma\delta})$.
Nevertheless, less interest has been noticed to more complex gravity
theories. It is worthy to mention that modification in scalar
curvature is useful in many ways. When one take the case of low
curvature, accelerating cosmic expansion can be observed \cite{ya8},
while the high curvature can be used to smoothen the singularities.
In this respect, Harko \textit{et al.} \cite{ya9} put forward the basis of
$f(R)$ gravity and gave the notion of $f(R,T)$ theory (where
quantity $T$ is induced by quantum effects or exotic imperfect
matter distributions) in which he made matter geometry coupling.
They solved dynamical equations interpreting some cosmological and
astronomical backgrounds by taking various $f(R,T)$ models.

Houndjo \cite{ya10} performed cosmological reconstruction in
$f(R,T)$ gravity and claimed that his models could possibility unify
cosmic accelerated and matter dominated eras. Jamil \emph{et al.}
\cite{ya11} discussed the reconstruction of some well-known
astrophysical models with $f(R,T)$ corrections and obtained results
consistent with low red-shifts Baryonic Acoustic Oscillations
observations. Adhav \cite{ya12} investigated exact solutions of some
cosmological models by taking exponential volumetric expansion in
this theory. Baffou \textit{et al.} \cite{ya13} investigated
dynamical evolution along with stability of power law and de-Sitter
cosmic models against linear perturbation. They concluded that such
models can be considered as a competitive dark energy candidate. Sun
and Huang \cite{ya14} addressed some cosmic issues of isotropic and
homogeneous universe in $f(R,T)$ gravity and found results
consistent with astronomical observation data.

Anisotropic effects are leading paradigms in addressing the
evolutionary mechanisms of celestial imploding models. The
assumption of considering isotropic nature of pressure distribution
in self-gravitating relativistic bodies is often under discussion by
many researchers. However, there are several arguments indicating
that the relativistic fluid pressure can be slightly varied in
different directions (anisotropic) at any particular point. Bowers
and Liang \cite{ya15} did pioneer work in describing possible
significance of locally anisotropic pressure distribution in
relativistic spherical matter configurations and found that
anisotropy may have worthwhile effects on parameters controlling the
hydrostatic equilibrium of celestial systems. The anisotropy within
stellar systems can be observed through number of interested
interconnected mechanisms, e.g., existence of strong electric and
magnetic interactions \cite{ya16}, condensations of pions
\cite{ya17}, phase transitions \cite{ya18}, the presence of vacuum
core \cite{ya19}, even the emergence of gravitational waves from
non-static meridional axial stellar structures \cite{ya20} etc. It
can be demonstrated that the mixture of two fluid configurations can
be mathematically treated as an anisotropic framework. Chakraborty
\textit{et al.} \cite{ya21} investigated pressure anisotropy
contributions on the collapsing quasi-spherical model and found that
such configurations of pressure could obstruct appearance of naked
singularity.
The dynamical analysis of a collapsing relativistic
stellar system has been performed~\cite{Arbuzova:2010iu, ya21a} and
it has been shown that the invoking of $R^\alpha(1<\alpha\leq2)$ corrections
could lead to a viable and singularity free model~\cite{ya21a}.

The characterization of gravitational collapse of stellar interiors
under numerous scenarios remain an open research window in
relativistic astrophysics. Ghosh and Maharaj \cite{ya22}
investigated dynamics of collapsing dust cloud with $f(R)$
corrections and found relatively stable fluid configurations against
perturbation mode. Cembranos \textit{et al.} \cite{ya23} studied
collapsing mechanism of relativistic dust particles and found highly
contracted configurations of collapsing systems due to the presence
of $f(R)$ gravitational interaction. Capozziello \textit{et al.}
\cite{ya24} calculated modified versions of Poisson and Boltzmann
dynamical equations for relativistic self-gravitating structures in
$f(R)$ gravity and found that some more unstable modes of the
evolving systems at N region due to $f(R)$ dark source terms.
Sebastiani \textit{et al.} \cite{ya25} described evolving phases of
spherical relativistic systems with $f(R)$ background and found a
wide range of different instability regions for the evolving compact
systems. Recently, Yousaf and Bhatti \cite{yb1} explored that some
$f(R)$ model configurations would support more compact cylindrical
objects with smaller radii as compared to general relativity (GR).

The dynamics of self-gravitating stellar systems can be addressed
with the help of system's structural variables, such as local
pressure anisotropy, energy density, Weyl scalar, etc. The density
irregularities and anisotropy occupy major character in the
collapsing mechanism and thus in developing theory of cosmic
structure formation. Any relativistic system begins collapsing once
it enters into an inhomogeneous state. Thus, in order to study
subsequent evolution of collapsing fluid configurations, one
requires to explore factors responsible for producing energy density
irregularities. In this perspective, Penrose and Hawking \cite{ya26}
explored irregularities in the energy density of spherical
relativistic stars by means of Weyl invariant. Herrera \textit{et
al.} \cite{ya27} evaluated inhomogeneity parameters for anisotropic
spherical compact objects and found that local anisotropy may yield
appearance of naked singularities. Mena \textit{et al.} \cite{ya28}
analyzed role of shearing and regular appearance of the collapsing
dust fluid on the subsequent evolution. Herrera \textit{et al.}
\cite{ya29} related Weyl scalar with fluid parameters and discussed
gravitational arrow of time for dissipative spherical star. Herrera
\textit{et al.} \cite{ya30} explored role of cosmological constant
in the irregularity factors and shear and evolution equations.
Sharif and his collaborators \cite{ya31} explored some inhomogeneity
and dynamical factors for tilted charged, conformally flat and
non-tilted relativistic systems with different backgrounds.
Recently, Sharif and Yousaf \cite{ya31a} found energy density
irregularity parameters in the subsequent evolution of celestial
objects in $f(R)$ gravity.

This paper extended the work \cite{ya30} in order to describe the
effects of modification of Einstein gravity in the formulation of
structure scalars. We also explore the role of these scalar
variables in the evolution equations for dissipative
self-gravitating spherical compact stars. The paper is organized as
follows. In section \textbf{2}, we describe spherical dissipative
matter configuration with $f(R,T)$ formalism and then make a
connection between structural variables with the Weyl invariant.
Section \textbf{3} is devoted to formulate modified scalar functions
and elaborate their role in the dynamics of self-gravitating systems
while section \textbf{4} discusses the contribution of these
structure variables for dust fluid with constant Ricci and trace of
stress-energy tensors. The main findings are concluded in the last
section.

\section{Spherical Dissipative Fluid Description with $f(R,T)$ Formalism}

We consider $f(R,T)$ gravity achieved by generalizing the action of
general relativity coupled with ordinary matter Lagrangian $L_M$ as \cite{ya9}
\begin{equation}\label{1}
S_{f(R,T)}=\int d^4x\sqrt{-g}[f(R,T)+L_M],
\end{equation}
where $g,~T$ are the traces of metric and usual energy-momentum
tensors, respectively while $R$ and $L_M$ represent Ricci scalar and
matter Lagrangian density. We chose the unit system, $8{\pi}G=c=1$.
The usual energy-momentum tensor can be found as
\begin{equation}\label{1n}
T_{\alpha\beta}=-\frac{2}{\sqrt{-g}}\frac{\delta(\sqrt{-g}L_M)}{\delta
g^{\alpha\beta}}.
\end{equation}
If we assume that $L_M$ depends merely on metric variables, i.e.,
$g_{\alpha\beta}$ (not upon its derivatives), then we have the
following form of energy momentum tensor
\begin{equation}\label{2n}
T_{\alpha\beta}=g_{\alpha\beta}L_M-2\frac{\partial(L_M)}{\partial
g^{\alpha\beta}}.
\end{equation}
Upon varying the modified EH action with respect to
$g^{\alpha\beta}$, we get
\begin{eqnarray}\label{3n}
&&
\delta S_{f(R,T)}=\int
\left\{f_R\delta(g^{\alpha\beta}R_{\alpha\beta})-\frac{f}{2}g_{\alpha\beta}\delta
g^{\alpha\beta}+f_T \frac{\delta T}{\delta g^{\alpha\beta}} \delta
g^{\alpha\beta}
\right. \nonumber\\ 
&& \left.
\hspace{28mm}
{}+\frac{1}{\sqrt{-g}}\frac{\delta(\sqrt{-g}L_M)}{\delta
g^{\alpha\beta}}\right\}\sqrt{-g}d^4x,
\end{eqnarray}
where subscripts $T$ and $R$ describe $\frac{\partial}{\partial T}$
and $\frac{\partial}{\partial R}$ operators, respectively.
Considering variations of Ricci scalar and Christoffel symbols, the
above equation can be recast as
\begin{align}\nonumber
\delta S_{f(R,T)}&=\int \left\{f_R g_{\alpha\beta}\Box\delta
g^{\alpha\beta}-f_R\nabla_\alpha\nabla_\beta\delta
g^{\alpha\beta}+f_T\frac{\delta(g^{\mu\nu}T_{\mu\nu})}{\delta
g^{\alpha\beta}}\delta
g^{\alpha\beta}-\frac{f}{2}g_{\alpha\beta}\delta
g^{\alpha\beta}\right.\\\label{4n} &\left.+f_RR_{\alpha\beta}\delta
g^{\alpha\beta}+\frac{1}{\sqrt{-g}}\frac{\delta(\sqrt{-g}L_M)}{\delta
g^{\alpha\beta}}\right\}\sqrt{-g}d^4x,
\end{align}
where $\nabla_\alpha$ represents covariant derivation while $\Box$
indicates $\nabla_\alpha\nabla^\alpha$ operator. Now, we consider
$T\equiv g^{\mu\nu}T_{\mu\nu}$ variations with respect to
$g^{\alpha\beta}$ as
\begin{align}\label{5n}
\frac{\delta(g^{\mu\nu}T_{\mu\nu})}{\delta
g^{\alpha\beta}}=T_{\alpha\beta}+\Theta^1_{\alpha\beta},
\end{align}
where
\begin{align}\label{6n}
\Theta^1_{\alpha\beta}\equiv g^{\mu\nu}\frac{\delta
T_{\mu\nu}}{\delta g^{\alpha\beta}}.
\end{align}
Keeping in mind partial integration of first and second terms of
Eq.(\ref{4n}), one can obtain the following configurations of
$f(R,T)$ field equation as
\begin{align}\label{7n}
R_{\alpha\beta}f_R-(\nabla_{\alpha}\nabla_{\beta}-g_{\alpha\beta}\Box)
f_R-\frac{f}{2}g_{\alpha\beta}=(1-f_T)T^{(m)}_{\alpha\beta}-f_T\Theta^1_{\alpha\beta}.
\end{align}
Now, we can continue our calculation after substituting the value of
$\Theta^1_{\alpha\beta}$ and this is possible once we have matter
Lagrangian. Variation of Eq.(\ref{2n}) provides
\begin{align}\label{8n}
\frac{\delta T_{\mu\nu}}{\delta g^{\alpha\beta}}=L_M\frac{\delta
g_{{\mu\nu}}}{\delta
g^{\alpha\beta}}+\frac{L_M}{2}g_{\alpha\beta}g_{\mu\nu}
-\frac{g_{\mu\nu}}{2}T_{\alpha\beta}-2\frac{\partial^2L_M}{\partial
g^{\alpha\beta}\partial g^{\mu\nu}}.
\end{align}
Using this relation in Eq.(\ref{6n}), we obtain
\begin{align}\label{9n}
\Theta^1_{\alpha\beta}=L_Mg_{\alpha\beta}-2T_{\alpha\beta}
-2\frac{\partial^2L_M}{\partial g^{\alpha\beta}\partial
g^{\mu\nu}}g^{\mu\nu}.
\end{align}
The choice of matter Lagrangian is directly connected with the value
of $\Theta^1_{\alpha\beta}$. As the dynamical equations in this
theory depends upon contribution from matter contents, therefore one
can obtain particular scheme of equations corresponding to every
selection of $L_M$. For example, for electromagnetic field theory
one can take $L_M=-F_{\mu\nu}F_{\zeta\eta}g^{\mu\zeta}g^{\nu\eta}$,
(where $F_{\mu\nu}$ is the Maxwell tensor) for which
$\Theta^1_{\alpha\beta}=-T_{\alpha\beta}$. Here, we are considering
the even more complex problem in which non-static geometry of
spherical system is coupled with shearing viscous and locally
anisotropic fluid configurations, radiating through heat flux and
free streaming approximation. We assume the following mathematical
expression of the  stress-energy tensor (along with $L_M=-\mu$)
\begin{equation}\label{4}
T_{\alpha\beta}=P_{\bot}h_{\alpha\beta}+{\mu}V_\alpha V_\beta+\Pi
\chi_\alpha\chi_\beta+{\varepsilon}l_\alpha l_\beta+q(\chi_\beta
V_\alpha+\chi_\alpha V_\beta)-2{\eta}{\sigma}_{\alpha\beta},
\end{equation}
where $\mu$ is the energy density, $q$ is a scalar quantity
corresponding to a heat conducting vector, $q_{\beta}$. The quantity
$q_\beta$ can be expressed by means of radial unit four vector,
$\chi_{\beta}={H}\delta_{\beta}^{1}$, as
$$q_\beta=q\chi_{\beta}.$$
Further, $\eta$ is the coefficient of shear viscosity, while
$\epsilon$ and $\sigma_{\alpha\beta}$ are radiation density and
shear tensor, respectively. Moreover,
$h_{\alpha\beta}=g_{\alpha\beta}+V_{\alpha}V_{\beta}$ is the
projection tensor and $\Pi$ is the difference of radial, $P_r,$ and
tangential pressure, $P_{\perp}$, given by $\Pi\equiv P_r-P_\bot$.
Now with the help of Eq.(\ref{9n}), we obtain
\begin{align}\label{9n}
\Theta^1_{\alpha\beta}=-2T_{\alpha\beta} -\mu g_{\alpha\beta}.
\end{align}
The corresponding $f(R,T)$ field equations are given as follows
\begin{equation}\label{2}
{G}_{\alpha\beta}={{T}_{\alpha\beta}}^{\textrm{eff}},
\end{equation}
where
\begin{align*}\nonumber
{{T}_{\alpha\beta}}^{\textrm{eff}}&=\left[(1+f_T(R,T))T_{\alpha\beta}+\mu
g_{\alpha\beta}f_T(R,T)
+\left(\frac{f(R,T)}{R}-f_R(R,T)\right)\frac{R}{2}g_{\alpha\beta}\right.\\\nonumber
&\left.+\left({\nabla}_\alpha{\nabla}_
\beta-g_{\alpha\beta}{\Box}\right)f_R(R,T)\right]\frac{1}{f_R(R,T)}
\end{align*}
is the effective energy-momentum tensor encapsulating gravitational
contribution coming from $f(R,T)$ extra degrees of freedom while
${G}_{\alpha\beta}$ is the Einstein tensor.

We consider non-static geometry of spherical system
\begin{equation}\label{3}
ds^2=-A^2(t,r)dt^{2}+H^2(t,r)dr^{2}+C^2d\theta^{2}
+C^2\sin^2\theta{d\phi^2},
\end{equation}
where $A,~H$ are dimension-less quantities wile $C$ has $L$
dimension. The quantities $V^{\beta}$ and $l^\beta$ in Eq.(\ref{4})
are fluid four-velocity and the null four-vector, respectively. The
four-vectors $V^{\beta}=\frac{1}{A}\delta^{\beta}_{0},~
\chi^{\beta},~
l^\beta=\frac{1}{A}\delta^{\beta}_{0}+\frac{1}{H}\delta^{\beta}_{1},$
and $q^\beta=q(t,r)\chi^{\beta}$ under co-moving coordinates obey
\begin{eqnarray*}
&&V^{\alpha}V_{\alpha}=-1,\quad\chi^{\alpha}\chi_{\alpha}=1,
\quad\chi^{\alpha}V_{\alpha}=0,\\\nonumber
&&V^\alpha q_\alpha=0, \quad l^\alpha V_\alpha=-1, \quad l^\alpha
l_\alpha=0.
\end{eqnarray*}
The kinematical scalars representing expansion and shearing motion
of spherical symmetric metric are given, respectively, as follows
\begin{equation*}
\Theta=\frac{1}{A}\left(\frac{2\dot{C}}{C}+\frac{\dot{H}}{H}\right),
\quad \sigma=\frac{1}{A}\left(\frac{\dot{H}}{H}
-\frac{\dot{C}}{C}\right),
\end{equation*}
where the over dot represents $\frac{\partial}{\partial t}$ operation.

The $f(R,T)$ field equations for spherically relativistic interior
system (with signatures $(-1,1,1,1)$)  are
\begin{align}\label{5}
G_{00}&=\frac{A^2}{f_{R}}\left[{\mu}+{\varepsilon}
-\frac{R}{2}\left(\frac{f}{R}-f_{R}\right)+\frac{\psi_{00}}{A^2}
\right],\\\label{6}
G_{01}&=\frac{AH}{f_{R}}\left[-(1+f_T)(q+{\varepsilon})
+\frac{\psi_{01}}{AH}\right],\\\label{7}
G_{11}&=\frac{H^2}{f_{R}}\left[\mu
f_T+(1+f_T)(P_r+\varepsilon-\frac{4}{3}\eta{\sigma})
+\frac{R}{2}\left(\frac{f}{R}-f_{R}\right)+\frac{\psi_{11}}{H^2}\right],\\\label{8}
G_{22}&=\frac{C^2}{f_{R}}\left[(1+f_T)({P_{\bot}}+\frac{2}{3}\eta{\sigma})+\mu
f_T
+\frac{R}{2}\left(\frac{f}{R}-f_{R}\right)+\frac{\psi_{22}}{C^2}\right],
\end{align}
where
\begin{align}\nonumber
\psi_{00}&=2\partial_{tt}f_R+\left(\frac{\dot{H}}{H}-2\frac{\dot{A}}{A}
+2\frac{\dot{C}}{C}\right)\partial_tf_R+\left(A^2\frac{H'}{H}-2AA'
-2A^2\frac{C'}{C}\right)\frac{\partial_rf_R}{H^2},\\\nonumber
\psi_{01}&=\partial_t\partial_rf_R-\frac{A'}{A}\partial_tf_R
-\frac{\dot{H}}{H}\partial_rf_R,\\\nonumber
\psi_{11}&=\partial_{rr}f_R-\frac{H^2}{A^2}\partial_{tt}f_R
+\left(H^2\frac{\dot{A}}{A}-2H^2\frac{\dot{C}}{C}-2H\dot{H}
\right)\frac{\partial_tf_R}{A^2} \\\nonumber 
&+\left(\frac{A'}{A}+2\frac{C'}{C}
-2\frac{H'}{H}\right)\partial_rf_R,\\\nonumber
\psi_{22}&=-C^2\frac{\partial_{tt}f_R}{A^2}+\frac{C^2}{A^2}\left(
\frac{\dot{A}}{A}-3\frac{\dot{C}}{C}-\frac{\dot{H}}{H}\right)
\partial_tf_R+\frac{C^2}{H^2}\left(\frac{C'}{C}+\frac{A'}{A}
-\frac{H'}{H}\right)\partial_rf_R.
\end{align}
Here, the prime indicates $\frac{\partial}{\partial r}$ operation.

The four-velocity of the relativistic collapsing fluid, $U$, can be
obtained by taking variations of areal radius of spherical systems
with its proper time as follows
\begin{eqnarray}\label{9}
U=D_{T}C=\frac{\dot{C}}{A}.
\end{eqnarray}
The Misner-Sharp mass function $m(t,r)$ is given by \cite{ya32}
\begin{equation}\label{10}
m(t,r)=\frac{C}{2}\left(1+\frac{\dot{C}^2}{A^2}
-\frac{C'^2}{H^2}\right).
\end{equation}
By making use of using Eqs.~(\ref{5})--(\ref{7}), (\ref{9}) and
(\ref{10}), the variation of spherical mass function with respect to
time and radius can be given, respectively, as follows
\begin{align}\label{11}
D_T{m}&=\frac{-1}{2f_{R}}\left[U\left\{(1+f_T)(\bar{P}_r-\frac{4}{3}\eta\sigma)+\mu
f_T+\frac{R}{2}\left(\frac{f}{R}-f_{R}\right)+\frac{\psi_{11}}{H^2}\right\}\right.\\\nonumber
&\left.+E
\left\{(1+f_T)\bar{q}-\frac{\psi_{01}}{AH}\right\}C^2\right],\\\label{12}
D_Cm&=\frac{C^2}{2f_R}\left[\bar{\mu}-\frac{R}{2}\left(\frac{f}{R}-f_{R}\right)
+\frac{\psi_{00}}{A^2}+\frac{U}{E}\left\{(1+f_T)\bar{q}
-\frac{\psi_{01}}{AH}\right\}\right].
\end{align}
where $D_{C}=\frac{1}{C'} \frac{\partial}{\partial r}$,
$\bar{P}_r=P_r+\varepsilon,~\bar{\mu}=\mu+\varepsilon$ and
$\bar{q}=q+\varepsilon$. The Integration of Eq.(\ref{12}) gives
\begin{align}\label{13}
m&=\frac{1}{2}\int^C_{0}\frac{C^2}{f_R}\left[\bar{\mu}-\frac{R}{2}
\left(\frac{f}{R}-f_{R}\right)
+\frac{\psi_{00}}{A^2}+\frac{U}{E}\left\{\frac{(1+f_T)}{f_R}\bar{q}
-\frac{\psi_{01}}{AH}\right\}\right]dC,
\end{align}
where $E\equiv \frac{C'}{H}$. This can be expressed with the help of
Eq.(\ref{9}) as
\begin{eqnarray}\label{14}
E\equiv\frac{C'}{H}=\left[1+U^{2}-\frac{2m(t,r)}{C}\right]^{1/2}.
\end{eqnarray}
This specific combinations of dissipation structural variables,
energy density and $f(R,T)$ corrections through mass function can be
achieved from Eq.(\ref{13}) and is given as
\begin{equation}\label{15}
\frac{3m}{C^3}=\frac{3\kappa}{2C^3}\int^r_{0}\left[\bar{\mu}
-\frac{R}{2}\left(\frac{f}{R}-f_{R}\right)
+\frac{\psi_{00}}{A^2}+\frac{U}{E}\left\{\frac{(1+f_T)}{f_R}\bar{q}
-\frac{\psi_{01}}{AH}\right\}C^2C'\right]dr.
\end{equation}
The well-known couple of components of the Weyl tensor are defined
as
\begin{equation*}
E_{\alpha\beta}=C_{\alpha\phi\beta \varphi}V^{\phi}V^{\varphi},\quad
H_{\alpha\beta}=\tilde{C}_{\alpha
\gamma\beta\delta}V^{\gamma}V^\delta= \frac{1}{2}\epsilon_{\alpha
\gamma \eta\delta}C^{\eta\delta}_{~~\beta{\rho}}V^\gamma V^{\rho},
\end{equation*}
where
$\epsilon_{\alpha\beta\gamma\delta}\equiv\sqrt{-g}\eta_{\alpha\beta\gamma\delta}$
with $\eta_{\alpha\beta\gamma\delta}$ is a Levi-Civita symbol, while
$E_{\alpha\beta}$ and $H_{\alpha\beta}$ represent electric and
magnetic Weyl tensor components, respectively. The electric
component, $E_{\alpha\beta}$ in view of unit four velocity and four
vectors can be given by
\begin{equation*}\nonumber
E_{\alpha\beta}=\mathcal{E}\left[\chi_{\alpha}\chi_{\beta}-\frac{1}{3}
(g_{\alpha\beta}+V_\alpha V_\beta)\right],
\end{equation*}
where
\begin{eqnarray}\nonumber
\mathcal{E}&=&\left[\frac{\ddot{C}}{C}+\left(\frac{\dot{H}}{H}-\frac{\dot{C}}{C}\right)
\left(\frac{\dot{C}}{C}+\frac{\dot{A}}{A}\right)-\frac{\ddot{H}}{H}\right]
\frac{1}{2A^{2}}-\frac{1}{2C^{2}}\\\label{16}
&-&\left[\frac{C''}{C}-\left(\frac{C'}{C}+\frac{H'}{H}\right)
\left(\frac{A'}{A}-\frac{C'}{C}\right)-\frac{A''}{A}\right]\frac{1}{2H^{2}}
\end{eqnarray}
is the Weyl scalar. This scalar after using Eqs.(\ref{10}) and
(\ref{15}) can be written as
\begin{align}\nonumber
\mathcal{E}&=\frac{1}{2f_R}\left[\bar{\mu}-(1+f_T)(\bar{\Pi}-2\eta\sigma)
-\frac{R}{2}\left(\frac{f}{R}-f_{R}\right)
+\frac{\psi_{00}}{A^2}-\frac{\psi_{11}}{H^2}
+\frac{\psi_{22}}{C^2}\right]\\\nonumber &-\frac{3}{2C^3}
\int^r_{0}\frac{C^2}{f_{R}}
\left[\bar{\mu}-\frac{R}{2}\left(\frac{f}{R}-f_{R}\right)
+\frac{\psi_{00}}{A^2}+\frac{U}{E}\left\{\frac{(1+f_T)}{f_R}\bar{q}
-\frac{\psi_{01}}{AH}\right\}\right.\\\label{17} &\left.\times
C^2C'\right]dr,
\end{align}
where $\bar{\Pi}=\bar{P}_r-P_\perp$. Here we have assumed a regular
matter configuration at the center, i.e., $m(t,0)=0=C(t,0)$. The
above expression provides a link between Weyl scalar, $f(R,T)$
higher curvature quantities and structural variables of matter
distribution (pressure anisotropy, radiation and energy density,
heat radiating vector, shearing viscosity).

\section{Modified Scalar Variables and $f(R,T)$ Gravity}

In this section, we firstly take a viable configuration of $f(R,T)$
model and then construct scalar functions after orthogonally
splitting Reimman tensor. In order to present $f(R,T)$ gravity as an
acceptable theory, one should consider viable as well as
well-consistent $f(R,T)$ models. Thus, we investigate the dynamical
properties of dissipating anisotropic spherical fluid distribution
by taking following configuration of $f(R,T)$ model \cite{ya33}
\begin{equation}\label{18}
f(R,T)=f_1(R)+f_2(T).
\end{equation}
This choice yields a minimal matter curvature coupling, thereby
presenting $f(R,T)$ gravity as corrections of $f(R)$ gravity.
Starobinsky \cite{ya7} suggested that quadratic Ricci scalar
corrections, i.e., $f(R)=R+\alpha R^2$ in the field equations could
be helpful to cause exponential early universe expansion. Several
relativisticts \cite{u30} adopted this formulation not only for an
inflationary constitute but also as a substitute for dark matter
(DM) for $\alpha=\frac{1}{6M^2}$ \cite{u31}. For DM model, $M$ is
figured out as $2.7\times10^{-12}\mathrm{GeV}$ with
$\alpha\leq2.3\times10^{22}\mathrm{GeV}^{-2}$ \cite{u19a}. This is the
minimum value of $M$ which follows from Cavendish-type laboratory
tests of the Newton law of gravity. It is interesting to mention
that extension to this model could provide a platform different from
that of $R+\alpha R^2$ to understand various cosmic puzzles. Here,
we take $f_1(R)=R+{\alpha}R^n-{\beta}R^{2-n}$ \cite{ya34} along with
$f_2(T)={\lambda}T$, with $\alpha,~\beta$ and ${\lambda}$ as
positive real numbers. The particular selection of $f_1(R)$ could be
constructive to discuss inflation from the $\alpha R^n$ along with a
stable minimum of the scalar potential of an auxiliary field. This
also assists to obtain a potential having a non-zero residual vacuum
energy, thereby providing it as a DE in the late-time cosmic
evolution.

In order to present acceptable $f(R,T)$ theory of gravity, one
should consider viable as well as well-consistent $f(R,T)$ models. A
viable model not only helps to shed light over current cosmic
acceleration but also obeys the requirements imposed by terrestrial
and solar system experiments with relativistic background. Further,
they should satisfy minimal constraints for theoretical viability.
Any modified gravity model needs to possess exact cosmological
dynamics and avoids the instabilities such as ghosts
(Dolgov-Kawasaki instability, Ostrogradski's instability and
tachyons). The following conditions should be satisfied for a viable
$f_1(R)$ models \cite{1c1}:
\begin{itemize}
\item The positive value of $f_{1R}(R)$ with $R>\tilde{R}$, here $\tilde{R}$
is the today value of the Ricci invariant. This condition is needed
to avoid appearance of a ghost state. Ghost often appears, while
dealing with modified gravity theories that informs DE as a source
behind current cosmic acceleration. This may be induced due to a
mysterious force which is repulsive in nature between massive or
supermassive stellar objects at large distances. The constraint of
keeping effective gravitational constant, $G_{eff}=G/f_{1R}$, to be
positive is also significant to retain the attractive feature of
gravity.
\item The positive value of $f_{1RR}(R)$ with $R>\tilde{R}$. This
condition is introduced to avoid the emergence of tachyons. A
tachyon is any hypothetical particle traveling faster than speed of
light. The moving mass of these particles would be imaginary and one
can assume the imaginary rest-mass so that moving mass must be real.
\end{itemize}
If $f_1(R)$ models do not satisfy these conditions, then it would be
regarded as unviable. Haghani \emph{et al.} \cite{yan1} and Odintsov
and Saez-Gomez \cite{yan2} suggested that Dolgov-Kawasaki
instability in $f(R,T)$ gravity requires similar sort of limitations
as in $f(R)$ gravity and in addition to this, we require $1+f_T>0$
for $G_{eff}>0$. So, for the viable $f(R,T)$ models, one needs to
satisfy the constraints
\begin{align}\nonumber
f_R>0,~1+f_T>0,~f_{RR}>0,~R\geq\tilde{R}.
\end{align}

It is worthy to stress that in our chosen $f(R,T)$ form, the term
$1+\lambda$ is explicitly taken to be positive. One thing that needs
to emphasized here is that the divergence of energy-momentum tensor
is non-zero in $f(R,T)$ gravity (unlike GR) and is found as
\begin{align}\label{14n}
\nabla^{\alpha}T_{\alpha\beta}=\frac{f_T}{1-f_T}\left[
(\Theta_{\alpha\beta}+T_{\alpha\beta})\nabla^\alpha{\ln}f_T
-\frac{1}{2}g_{\alpha\beta}\nabla^\alpha{T}
+\nabla^\alpha\Theta_{\alpha\beta}\right].
\end{align}
This makes breaking of all equivalence principles in $f(R,T)$
gravity. The weak equivalence principle states ``\emph{All test
particles in a given gravitational field, will undergo the same
acceleration, independent of their properties, including their rest
mass}". But in $f(R,T)$ gravity, the equation of motion depends on
the thermodynamic properties of the particles (energy density and
pressure etc). The strong equivalence principle, ``\emph{The
gravitational motion of a small test body depends only on its
initial position and velocity, and not on its constitution}" is
again obviously broken because of the non-geodesic motion of the
particles. The dynamics of $f(R)$ gravity can be recovered, on
setting $f(T)=0$.

Following Bel \cite{ya35} and Herrera \textit{et al.}
\cite{ya29,ya30}, we introduce couple of following tensors
$Y_{\alpha\beta}$ and $X_{\alpha\beta}$ as
\begin{equation*}
Y_{\alpha\beta}=R_{\alpha\gamma\beta\delta}V^{\gamma}V^{\delta},\quad
X_{\alpha\beta}=^{*}R^{*}_{\alpha\gamma\beta\delta}V^{\gamma}V^{\delta}=
\frac{1}{2}\eta^{\varepsilon\rho}_{~~\alpha\gamma}R^{*}_{\epsilon
\rho\beta\delta}V^{\gamma}V^{\delta},
\end{equation*}
where $R^{*}_{\alpha\beta\gamma\delta}=
\frac{1}{2}\eta_{\varepsilon\rho\gamma\delta}R^{\epsilon\rho}_{~~\alpha
\beta}$. In order to develop formalism for structure scalars in
$f(R,T)$ gravity, we orthogonally decomposed Riemann curvature
tensor. and found that
\begin{align}\nonumber
X_{\alpha\beta}&=X_{\alpha\beta}^{(m)}+X_{\alpha\beta}^{(D)}=\frac{1}{3f_R}\left[
\bar{\mu}-\frac{R}{2}\left(\frac{f}{R}-f_{R}\right)+\frac{\psi_{00}}{A^2
}\right]h_{\alpha\beta}\\\label{20a}
&-\frac{1}{2f_R}\left[(1+\lambda)(\bar{\Pi}-2\eta\sigma)+\frac{{\psi}_{11}}{H^2}
-\frac{{\psi}_{22}}{C^2}\right]\left(\chi_\alpha\chi_\beta-\frac{1}{3}h_{\alpha\beta}\right)-E_{\alpha\beta}\\\nonumber
Y_{\alpha\beta}&=Y_{\alpha\beta}^{(m)}+Y_{\alpha\beta}^{(D)}=\frac{1}{6f_R}\left[\bar{\mu}+3\mu\lambda+(1+\lambda)
(3P_r-2\bar{\Pi})+\frac{\psi_{00}}{A^2}+\frac{\psi_{11}}{H^2}\right.\\\nonumber
&\left.+\frac{2\psi_{22}}{C^2}+R\left(\frac{f}{R}-f_{R}\right)\right]h_{\alpha\beta}+\frac{1}{2f_R}
\left[(1+\lambda)(\bar{\Pi}-2\eta\sigma) +\frac{{\psi}_{11}}{H^2}
-\frac{{\psi}_{22}}{C^2}\right]\\\label{20b}
&\times\left(\chi_\alpha\chi_\beta
-\frac{1}{3}h_{\alpha\beta}\right)-E_{\alpha\beta}.
\end{align}
These tensors can be written as a combination of their trace and
trace-less components as follows
\begin{align}\label{19}
X_{\alpha\beta}&=\frac{1}{3}TrXh_{\alpha\beta}+X_{<\alpha\beta>},\\\label{20}
Y_{\alpha\beta}&=\frac{1}{3}TrYh_{\alpha\beta}+Y_{<\alpha\beta>},
\end{align}
where
\begin{align}\label{21}
X_{<\alpha\beta>}&=h^\rho_\alpha
h^\gamma_\beta\left(X_{\rho\gamma}-\frac{1}{3}TrXh_{\rho\gamma}\right),\\\label{22}
Y_{<\alpha\beta>}&=h^\rho_\alpha
h^\gamma_\beta\left(Y_{\rho\gamma}-\frac{1}{3}TrYh_{\rho\gamma}\right).
\end{align}
{}From Eqs.(\ref{18})-(\ref{20b}), we found
\begin{align}\nonumber
&TrX\equiv
X_{T}=\frac{1}{1+n{\alpha}R^{n-1}-\beta(2-n)R^{1-n}}\left\{\bar{\mu}
-\frac{\alpha(1-n)}{2}R^n+\frac{\beta(3-n)}{2}\right.\\\label{23}
&\times\left.R^{2-n}-\frac{{\lambda}}{2}T
+\frac{\hat{\psi}_{00}}{A^2}\right\},\\\nonumber &TrY\equiv
Y_{T}=\frac{1}{2(1+n{\alpha}R^{n-1}-\beta(2-n)R^{1-n})}\left\{\bar{\mu}+3\mu\lambda
+3(1+\lambda)\bar{P_r}\right.\\\label{24}
&\left.-2(1+\lambda)\bar{\Pi}+\frac{\hat{\psi}_{00}}{A^2}+\frac{\hat{\psi}_{11}}{H^2}+\frac{2\hat{\psi}_{22}}
{C^2}+2\alpha(1-n)R^n+2\beta(3-n) R^{2-n}-2{\lambda}T\right\},
\end{align}
where hat indicates that the corresponding dark source terms are
evaluated after using $f(R,T)$ model. We can also write
$X_{<\alpha\beta>}$ and $Y_{<\alpha\beta>}$ in an alternatively form
\begin{align}\label{25}
X_{<\alpha\beta>}&=X_{TF}\left( \chi_{\alpha}\chi_{\beta}
-\frac{1}{3}h_{\alpha\beta}\right),\\\label{26}
Y_{<\alpha\beta>}&=Y_{TF}\left( \chi_{\alpha}\chi_{\beta}
-\frac{1}{3}h_{\alpha\beta}\right),
\end{align}
where the quantities $X_{TF}$ and $X_{TF}$ are 
\begin{align}\nonumber
X_{TF}&=-\mathcal{E}-\frac{1}{2(1+n{\alpha}R^{n-1}-\beta(2-n)R^{1-n})}
\left\{(1+\lambda)(\bar{\Pi}-2{\sigma}{\eta})\right.\\\label{27}
&\left.+\frac{\hat{\psi}_{11}}{H^2}
-\frac{\hat{\psi}_{22}}{C^2}\right\},\\\label{28}
Y_{TF}&=\mathcal{E}-\frac{1}{2(1+n{\alpha}R^{n-1}-\beta(2-n)R^{1-n})}
\left\{(\bar{\Pi}-2{\eta}{\sigma})(1+\lambda)
\right. \nonumber \\
&\left.+\frac{\hat{\psi}_{11}}{H^2}
-\frac{\hat{\psi}_{22}}{C^2}\right\}.
\end{align}
The scalar function $Y_{TF}$ can be written in terms of matter
variables after using Eqs.(\ref{17}), (\ref{21}) and (\ref{24}) as
\begin{align}\nonumber
Y_{TF}&=\frac{1}{2(1+n{\alpha}R^{n-1}-\beta(2-n)R^{1-n})}\left(\bar{\mu}-2(1+\lambda)(\bar
{\Pi}-4{\eta}{\sigma})+\frac{\alpha}{2}\right.\\\nonumber
&\left.\times(1-n)R^n-\frac{\beta}{2}(3-n)R^{2-n}
+\frac{{\lambda}}{2}T+\frac{\hat{\psi}_{00}}{A^2}-\frac{2\hat{\psi}_{11}}{H^2}
+\frac{2\hat{\psi}_{22}}{C^2}\right)-\frac{3}{2C^3}\\\nonumber
&\times \int^r_{0}\frac{C^2}{1+n{\alpha}R^{n-1}-\beta(2-n)R^{1-n}}
\left[\bar{\mu}-\frac{\alpha}{2}(1-n)R^n+\frac{\beta}{2}(3-n)R^{2-n}
\right.\\\label{29}
&\left.-\frac{\lambda}{2}T+\frac{\hat{\psi}_{00}}{A^2}+\frac{U}{E}
\left\{\frac{(1+\lambda)}{1+n{\alpha}R^{n-1}-\beta(2-n)R^{1-n}}\bar{q}
-\frac{\hat{\psi}_{01}}{AH}\right\}C^2C'\right]dr.
\end{align}
Now we express fluid variables by defining some effective variables
\begin{align*}
\mu_{eff}&\equiv\bar{\mu}+\frac{\hat{\psi}_{00}}{A^2}, \quad
P^{eff}_{r}\equiv
\bar{P_r}+\frac{\hat{\psi}_{11}}{H^2}-\frac{4}{3}{\eta}{\sigma},\\
P^{eff}_{\bot}&\equiv P_{\bot}+\frac{\hat{\psi}_{22}}{C^2}+\frac{2}{3}{\eta}\sigma,\\
\Pi^{eff}&\equiv P^{eff}_{r}-P^{eff}_{\bot}=\Pi-2{\eta}{\sigma}
-\frac{\psi_{22}}{C^2}+\frac{\psi_{11}}{H^2}.
\end{align*}
These terms are just like the usual matter structure variables with
the difference that they have modified gravity as well as viscosity
terms in some specific combination. Using above effective variables,
Eqs.(\ref{23}), (\ref{24}), (\ref{27}) and (\ref{28}) reduce to
\begin{align}\nonumber
X_{TF}&=\frac{3\kappa}{2C^3}
\int^r_{0}\left[\frac{1}{\{1+n{\alpha}R^{n-1}-\beta(2-n)R^{1-n}\}}\left\{\mu_{eff}
-\frac{\alpha}{2}(1-n)R^n+\frac{\beta}{2}\right.\right.\\\nonumber
&\times\left.\left.(3-n)R^{(2-n)}-\frac{\lambda}{2} T
+\left(\hat{q}-\frac{\psi_q}{AB}\right)\frac{U}{E}\right
\}C^2C'\right]dr\\\nonumber
&-\frac{1}{2\{1+n{\alpha}R^{n-1}-\beta(2-n)R^{1-n}\}}\left[\mu_{eff}-\frac{\alpha}{2}(1
-n)R^n+\frac{\beta}{2}(3-n)\right.\\\label{30} &\left.\times
R^{(2-n)}-\frac{\lambda}{2} T\right],\\\nonumber
Y_{TF}&=\frac{1}{2(1+n{\alpha}R^{n-1}-\beta(2-n)R^{1-n})}\left[\mu_{eff}-
\frac{\alpha}{2}(1-n)R^n+\frac{\beta}{2}\right.\\\nonumber
&\times\left.(3-n)R^{(2-n)}-\frac{\lambda}{2} T-
2(1+\lambda)\Pi^{eff}+2\lambda\left(\frac{\hat{\psi}_{11}}{H^2}
-\frac{\hat{\psi}_{22}}{C^2}\right)\right]\\\nonumber
&-\frac{3}{2C^3}\int^r_{0}\left[\frac{1}{\{1+n{\alpha}R^{n-1}-\beta(2-n)R^{1-n}\}}
\left\{\mu_{eff}-\frac{\alpha}{2}(1-n)R^n\right.\right.\\\label{31}
&\left.\left.+\frac{\beta}{2}(3-n)R^{(2-n)}-\frac{\lambda}{2} T
+\left(\hat{q}-\frac{\psi_q}
{AH}\right)\frac{U}{E}\right\}C^2C'\right]dr,\\\nonumber
Y_{T}&=\frac{1}{2(1+n{\alpha}R^{n-1}-\beta(2-n)R^{1-n})}\left[(1+3\lambda)\mu_{eff}
-3\varepsilon\lambda+3(1+\lambda)\right.\\\nonumber &\left.\times
P_r^{eff}-2(1+\lambda)\Pi^{eff}-\lambda\left(\frac{\psi_{11}}{H^2}+3
\frac{\psi_{00}}{A^2}\right)+2(2+\lambda)\frac{\psi_{22}}{C^2}-2\alpha\right.\\\label{32}
&\left.\times(1-n)-2\beta(3-n)R^{(2-n)}-2\lambda T\right],
\\\nonumber
X_{T}&=\frac{1}{(1+n{\alpha}R^{n-1}-\beta(2-n)R^{1-n})}\left[\mu_{eff}-
\frac{\alpha}{2}(1-n)R^n+\frac{\beta}{2}(3-n)\right.\\\label{33}
&\left.\times R^{2-n}-\frac{{\lambda}}{2}T\right].
\end{align}
On setting $\lambda=0$, the $f(R)$ structure \cite{ya31a} scalars
can be retrieved from the above expressions. These structure
functions have a direct correspondence with the dynamical evolution
of relativistic compact systems even in $f(R,T)$ gravity theory. It
is evident from Eq.(\ref{33}) that $X_T$ has foremost importance in
the definition of stellar energy density along with the effects of
dark source terms coming from $f(R,T)$ gravity. Following Herrera
\textit{et al.} \cite{ya30}, the evolution equation connecting
effects of tidal forces with fluid parameters variables is
\begin{align}\nonumber
&\left[X_{TF}+\frac{1}{2\{1+n{\alpha}R^{n-1}-\beta(2-n)R^{1-n}\}}
\left\{\bar{\mu}_{eff}-\frac{\alpha(1-n)}{2}R^n+\frac{\beta}{2}(3-n)\right.\right.\\\nonumber
&\left.\left.\times R^{2-n}-\frac{{\lambda}}{2}T\right\}\right]'=
-X_{TF}\frac{3C'}{C}+\frac{(\sigma-\Theta)}{2[1+n{\alpha}R^{n-1}-\beta(2-n)R^{1-n}]}
\left[\frac{\psi_{01}}{H}\right.\\\label{34}
&\left.-\frac{(1+\lambda)}{[1+n{\alpha}R^{n-1}-\beta(2-n)R^{1-n}]}\bar{q}H\right].
\end{align}
It is evident from the above equation that in the absence of dark
source and radiating variables, one can obtain the following result
after considering regularity constraints as
\begin{align}\nonumber
\mu'_{eff}=0\Leftrightarrow X_{TF}=0.
\end{align}
This means that $X_{TF}$ controls the inhomogeneity of the
collapsing star. In the non-radiating isotropic matter distribution,
we can get from Eq.(\ref{34}), that $\mu'_{eff}$ only exists if and
only if $\mathcal{E}$ exists. This suggests that tidal forces try to
move the self-gravitating compact objects into inhomogeneous window
as the time proceeds. This led Penrose to describe a gravitational
time arrow with the help of the Weyl tensor in GR. To check the role
of rest of structure variables, we consider well-known mathematical
tool put forward through the so-called Raychaudhuri equation (also
calculated individually by Landau) \cite{ya36}. In view of one of
the modified structure scalars, it follows that
\begin{equation}\label{35}
-Y_{T}=V^{\alpha}\Theta_{;\alpha}+\frac{2}{3}{\sigma}^{
\alpha\beta}{\sigma}_{\alpha\beta}+\frac{{\Theta}^{2}}{3}
-a^\alpha_{~;\alpha}.
\end{equation}
This relation shows that $Y_T$ has an utmost relevance in the
description of the expansion rate of self-gravitating relativistic
fluids. The equation describing the shear evolution can be recast in
terms of $Y_{TF}$ as follows
\begin{equation}\label{36}
Y_{TF}=a^{2}+\chi^{\alpha}a_{;\alpha}-\frac{aC'}{HC}
-\frac{2}{3}{\Theta}\sigma-V^\alpha
\sigma_{;\alpha}-\frac{1}{3}\sigma^{2},
\end{equation}
thus describing that $f(R,T)$ correction terms have its importance
in the shearing motion of the evolving relativistic spherical
self-gravitating system.

\section{Evolution Equations with Constant $R$ and $T$}

Here, we discuss the contribution of modified structure scalars for
the dust spherical cloud with Ricci scalar and $T\equiv
T^\beta_{~\beta}$ background. In this context, the quantity of
matter within the spherical model of radius $r$ is
\begin{align}\nonumber
m&=\frac{1}{2\{1+n{\alpha}\tilde{R}^{n-1}-\beta(2-n)\tilde{R}^{1-n}\}}
\int^r_{0}(\mu) C^2C'dr\\\label{37}
&-\frac{\alpha(1-n)\tilde{R}^n-\beta(3-n)\tilde{R}^{2-n}+{\lambda}\tilde{T}}
{2\{1+n{\alpha}\tilde{R}^{n-1}-\beta(2-n)\tilde{R}^{1-n}\}}\int^r_{0}
C^2C'dr,
\end{align}
where tilde indicates that the corresponding terms are evaluated
under constant backgrounds. For the dust interior cloud, the couple
of equations describing the tidal forces and peculiar form of mass
function can be given as follows
\begin{align}\nonumber
\mathcal{E}&=\frac{1}{2C^3\{1+n{\alpha}\tilde{R}^{n-1}-\beta(2-n)\tilde{R}^{1-n}\}}
\int^r_{0}\mu'C^3dr\\\label{38}
&-\frac{\alpha(1-n)\tilde{R}^n-\beta(3-n)\tilde{R}^{2-n}+{\lambda}\tilde{T}}
{4\{1+n{\alpha}\tilde{R}^{n-1}-\beta(2-n)\tilde{R}^{1-n}\}},\\\nonumber
\frac{3m}{C^3}&=\frac{1}{2\{1+n{\alpha}\tilde{R}^{n-1}-\beta(2-n)\tilde{R}^{1-n}\}}
\left[\mu-\frac{1}{C^3}\int^r_{0}\mu'C^3dr\right]\\\label{39}
&+\frac{\alpha(1-n)\tilde{R}^n-\beta(3-n)\tilde{R}^{2-n}+{\lambda}\tilde{T}}
{2\{1+n{\alpha}\tilde{R}^{n-1}-\beta(2-n)\tilde{R}^{1-n}\}}.
\end{align}
These equations are equivalent to Eqs.(\ref{15}) and (\ref{18}). The
$f(R,T)$ scalar functions in the realm of constant Ricci and $T$
turn out to be
\begin{align}\nonumber
\tilde{X}_{T}&=\frac{1}{\{1+n{\alpha}\tilde{R}^{n-1}-\beta(2-n)\tilde{R}^{1-n}\}}\left[{\mu}
-\frac{\alpha}{2}(1-n)\tilde{R}^n=\frac{\beta}{2}(3-n)\right.\\\label{40}
&\left.\times
\tilde{R}^{2-n}-\frac{{\lambda}}{2}\tilde{T}\right],\quad
\tilde{Y}_{TF}=-\tilde{X}_{TF}=\mathcal{E},\\\nonumber
\tilde{Y}_{T}&=\frac{1}{2\{1+n{\alpha}\tilde{R}^{n-1}-\beta(2-n)\tilde{R}^{1-n}\}}\left[{\mu}+3\mu\lambda
-2{\alpha}(1-n)\tilde{R}^n+2{\beta}\right.\\\label{41}
&\left.\times(3-n)\tilde{R}^{2-n}-2{{\lambda}}\tilde{T}\right],
\end{align}
where tilde indicates that the quantities are computed under
constant Ricci scalar condition. It is clear from Eq.(\ref{40}) that
$X_T$ describes matter energy density in the mysterious dark
universe while the evolution equation that describes the behavior of
the regular energy density over the relativistic dust cloud, can be
expressed by means of $X_{TF}$ as follows
\begin{align}\nonumber
&\left[\frac{\mu}{2\{1+n{\alpha}\tilde{R}^{n-1}-\beta(2-n)\tilde{R}^{1-n}\}}
-\frac{\alpha(1-n)\tilde{R}^n-\beta(3-n)\tilde{R}^{2-n}+{\lambda}\tilde{T}}{4\{1+n{\alpha}\tilde{R}^{n-1}
-\beta(2-n)\tilde{R}^{1-n}\}}\right.\\\label{42}
&\left.+\tilde{X}_{TF}\right]' =-\frac{3}{C}\tilde{X}_{TF}C'.
\end{align}
It follows from the above equation that the constant $f(R,T)$
corrections controls the irregularity in the energy density of the
matter distribution. The scenario $\mu'=0$ is directly related with
the vanishing of $\tilde{X}_{TF}$. This reinforces the importance of
structure scalar, $X_{TF}$ in the modeling of self-gravitating
matter configurations. The shear and expansion evolution equations
can be written in terms of rest of structure scalars as follows
\begin{align}\nonumber
&V^\alpha\Theta_{;\alpha}+\frac{2}{3}\sigma^2 +\frac{\Theta^2}{3}
-a^\alpha_{;\alpha}
\frac{1}{\{1+n{\alpha}\tilde{R}^{n-1}-\beta(2-n)\tilde{R}^{1-n}\}}\left[{\mu}
-\frac{\alpha}{2}(1-n)\right.\\\label{43} &\left.\times
\tilde{R}^n=\frac{\beta}{2}(3-n)
\tilde{R}^{2-n}-\frac{{\lambda}}{2}\tilde{T}\right]=-\tilde{Y}_T,\\\label{44}
&V^{\alpha}\sigma_{;\alpha}+\frac{\sigma^{2}}{3}+\frac{2}{3}\sigma
\Theta=-\mathcal{E}=-\tilde{Y}_{TF}.
\end{align}

\section{Conclusions}

In the present paper, we have discussed the dynamical properties of
compact objects by taking into account the well-known $f(R,T)$ high
degrees of freedom. First, we have studied spherical
self-gravitating system coupled with relativistic viscous matter
distribution that is radiating with free streaming out and diffusion
approximations. After evaluating the basic formulae, we have related
the system structural variables to the Weyl scalar. We have then
investigated factors that affect the contribution of tidal forces in
the evolution of collapsing spherical matter distribution in the
realm of $f(R,T)$ gravity. In order to bring out the effects of
modified gravity corrections, we have considered a particular class
of $f(R,T)$ models, i.e., the form is given by
$f(R,T)=f_1(R)+f_2(T)$. This choice does not imply the direct
non-minimal curvature matter coupling. Nevertheless, it can be
regarded as a correction to $f(R)$ gravity. We have used the linear
form of $f_2$ and acquired distinct results on the basis of a
non-trivial coupling in comparison with $f(R)$ gravity.

We have explored the role of $f(R,T)$ dark source terms in the
expressions of structure scalars. These scalars have been obtained
from the orthogonal decomposition of the Riemann curvature tensor.
We have found that as in general relativity,
these scalar variables controls the evolutionary mechanisms of radiating
fluid spheres in cosmos and they are four in number.

Our main results are summarized as follows.

(i)
It is found from Eq.(\ref{23}) that one of the structure scalar
(which is the trace part of Eq.(\ref{20a})) describes the energy
density (along dark source terms) of dissipative anisotropic
spherical distribution. This also shows that $f(R,T)$ correction
affects the contribution of $X_T$ due to its non-attractive nature.

(ii) It is well-known from the working of \cite{ya37} that structure
scalar, $Y_T$, has a direct link with Tolman mass ``density" for
dynamical systems to be in the phases of
equilibrium/quasi-equilibrium. We found that the role of this
quantity is controlled by anisotropic pressure along with dark
energy/matter terms. More appropriately, we can say that Tolman man
density is conspicuously related with pressure anisotropy, radiating
and non-radiating energy density along $f(R,T)$ correction. It is
seen from Eq.(\ref{24}) that even in radiating spheres, $Y_T$ can
have a direct link with non-dissipative energy density due to the
presence of $3\mu\lambda$ (this term comes due to $f(R,T)$ gravity).
Thus $f(R,T)$ gravity enhances the contribution of energy density in
the description of Tolman mass.

(iii)
The expansion scalar has utmost relevance with vacuum core emergence
within the stellar interior (see \cite{ya39}). Further, it is
well-known that expansion-free constraint requires pressure
anisotropy. It is seen from Eqs.~(\ref{35}) and (\ref{43}) that the
evolution of expansion scalar is fully controlled by $Y_T$.  Thus,
$Y_T$ may be helpful to understand the emergence of vacuum cavity
within the celestial object. This sparks that $Y_T$ should have a
direct relation with pressure anisotropy along with $f(R,T)$
corrections and this is obvious from Eqs.~(\ref{32}).

(iv) The structure scalar, $Y_{TF}$ depicts the influence of both
local pressure anisotropy, shear viscosity along with tidal forces
in the mysterious dark universe as seen from Eq.~(\ref{28}).
Furthermore, this scalar variable fully control the shear evolution
expression as mentioned by Eq.(\ref{36}). Thus, in order to
understand the role the shear motion on the dynamical phases of the
radiating celestial object, one needs to study the behavior of
$Y_{TF}$.

(v) Any celestial system must experience inhomogeneous state in
order to move into the collapsing phase. We found that the quantity
that controls irregularities in the energy density of stellar
interior is $X_{TF}$. It is well-known from the working of
\cite{ya29} that $X_{TF}$ behaves as an inhomogeneity factor for
dust perfect and anisotropic fluid. However, Eq.(\ref{34}) describes
that dissipative parameters as well as $f(R,T)$ corrections tend to
produce hindrances in the role of $X_{TF}$. However, if one
considers that expansion scalar is proportional to shear scalar,
then it is only $f(R,T)$ degrees of freedom that tends to produce
hindrances in the appearance of inhomogeneities in the relativistic
self-gravitating systems. Thus, $X_{TF}$ along with dark source
terms control density irregularities, and thus should be the basic
ingredient in the definition of a gravitational arrow of time.

(vi)
For constant curvature background with dust cloud matter distribution,
it is seen that it is the tidal forces (that are expressed with the help of
$X_{TF}$) which are responsible for producing irregularities in the
initial homogeneous stellar system.

(vii) All of our results are reduced to those in Ref.~\cite{ya30}
by taking $f(R,T)=R$.

Consequently, it has been shown that $f(R,T)$ gravity tends to
lessen down density homogeneity, which in turn induce the stability
to the relativistic collapsing compact systems. The $f(R,T)$ gravity
gives a new corrections to the EH Lagrangian through the coupling of
matter and geometry. In this theory, the cosmic acceleration depends
not only on a geometrical contribution to the total cosmic energy
density but also on the cosmic matter. If one takes
$f(R,T)=R+\lambda T$, then one can obtain dynamics identical with
that of GR. Thus, the simple case $f(R,T)=R+\lambda T$ is fully
equivalent with standard GR, after rescaling of $\lambda $.

\section*{Acknowledgments}

We would like to sincerely thank Professor Muhammad Sharif for his
kind encouragements on this study. This work was partially supported
by the JSPS Grant-in-Aid for Young Scientists (B) \# 25800136 and
the research-funds presented by Fukushima University (K.B.). This
work was also partially supported by University of the Punjab,
Lahore-Pakistan through research project in the fiscal year
2015-2016 (Z.Y.).

\end{document}